# Isospin Diffusion and Equilibration for Sn+Sn collisions at E/A=35 MeV

Z.Y. Sun(孙志宇)[1,2], M.B. Tsang(曾敏兒)[1,3], W.G. Lynch(連致標)[1,3], G. Verde[4], F. Amorini[5,6], L. Andronenko[1,7], M. Andronenko[1,7], G. Cardella[4], M. Chatterje[5], P. Danielewicz[1,3], E. De Filippo[4], P. Dinh[1], E. Galichet[5], E. Geraci[4], H. Hua (华辉)[8], E. La Guidara[4,9], G. Lanzalone[4,6], H. Liu(刘航)[10], F. Lu(卢飞)[3,8], S. Lukyanov[11], C. Maiolino[5], A. Pagano[4], S. Piantelli[12], M. Papa[4], S. Pirrone[4], G. Politi[4,6], F. Porto[5,6], F. Rizzo[5,6], P. Russotto[5,6], D. Santonocito[5], Y.X. Zhang(张英逊)[13]

[1]National Superconducting Cyclotron Laboratory and Department of Physics and Astronomy, Michigan State University, East Lansing, MI 48824, USA,
[2]Institute of Modern Physics, CAS, Lanzhou 730000, China,
[3]Joint Institute of Nuclear Astrophysics, Michigan State University, East Lansing, MI 48824, USA
[4]INFN, Sezione di Catania, Italy
[5]INFN, Laboratori Nazionali del Sud, Catania, Italy
[6]Departimento di Fisica e Astronomia, Universita di Catania, Italy,
[7]PNPI, Gatchina, Leningrad district 188300, Russian Federation.
[8]Peking University, Beijing, China,
[9]Centro Siciliano di Fisica Nucleare e Struttura della Materia, Catania, Italy
[10]University of Texas, Austin, TX 78705, USA
[11]FLNR, JINR, 141980 Dubna, Moscow region, Russian Federation
[12]INFN, Florence, Italy
[13]China Institute of Atomic Energy, Beijing, China,

## Abstract

Equilibration and equilibration rates have been measured by colliding Sn nuclei with different isospin asymmetries at beam energies of E/A=35 MeV. Using the yields of mirror nuclei of $^7$Li and $^7$Be, we have studied the diffusion of isospin asymmetry by combining data from asymmetric $^{112}$Sn+$^{124}$Sn and $^{124}$Sn+$^{112}$Sn collisions with that from symmetric $^{112}$Sn+$^{112}$Sn and $^{124}$Sn+$^{124}$Sn collisions. We use these measurements to probe isospin equilibration in central collisions where nucleon-nucleon collisions are strongly blocked by the Pauli exclusion principle. The results are consistent with transport theoretical calculations that predict a degree of transparency in these collisions, but inconsistent with the emission of intermediate mass fragments by a single chemically equilibrated source. Comparisons with ImQMD calculations are consistent with results obtained at higher incident energies that provide constraints on the density dependence of the symmetry energy.

The excitations of the various degrees of freedom during a central nucleus-nucleus collision evolve towards thermal equilibrium at different rates. For incident velocities exceeding the Fermi velocity, collective motions clearly do not become thermalized, but remain significant, providing insight into the reaction dynamics and the underlying nuclear equation of state [1, 2]. In contrast, particle emission and fragmentation appear to be thermalized in many respects, providing support for statistical interpretations that invoke complete chemical equilibrium of a "participant" source created by the overlap of projectile and target nucleons [3, 4]. Such interpretations appear to be inconsistent with transport theoretical interpretations of collisions at relative velocities that exceed the Fermi velocity, in which significant fractions of the projectile and target nucleons diffuse through each other without stopping and equilibrating [5, 6]. These studies have stimulated further investigations on how different degrees of freedom (thermal, mechanical, chemical) evolve during the collision and possibly achieve their freeze-out values at different times. In particular, the neutron/proton and N/Z-asymmetry of fragments have been recognized as one of the most quickly evolving degrees of freedom since the early studies of heavy-ion collisions at low and intermediate energies [7-13]. The hierarchical nature of heavy-ion collision dynamics suggests that the study of attainment of N/Z equilibrium may set important constraints on the assumption of global equilibrium commonly taken for granted in statistical theories of nuclear multifragmentation.

The study of N/Z diffusion and equilibration in heavy-ion collisions has recently received intense interests due to its links to transport properties of asymmetric nuclear matter [14-17]. When two nuclei with different N/Z asymmetries come into contact, diffusion of neutrons and protons is initiated and continues until the system disintegrates or until the chemical potentials for neutrons and protons in both nuclei become equal [18]. The rate of diffusion is influenced by the initial densities of neutrons and protons in the emitting nuclei, the neutron and proton mean free paths, and the mean field potentials, to which the symmetry energy contributes [14, 17, 18-21].

The number of diffused neutrons or protons is also sensitive to the total contact time. At low bombarding energies (E/A<20 MeV), characterized by relatively long interaction times between projectile and target, mid-peripheral reactions are dominated by deep inelastic collisions (DIC) processes where the diffusion of nucleons leads to an equilibration of the N/Z asymmetry over very short times scales [7, 10-13]. Microscopic calculations [12] have shown that already at low energies the attainment of isospin equilibration is driven by the strength of the symmetry potential due to the isovector term of the nucleon-nucleon interaction [12]. When the energy is increased around the Fermi energy domain, the time scale for fragmentation decay becomes comparable to or shorter than the time scales characterizing the attainment of isospin equilibration. Mirror nucleus yield ratios of $^7$Li and $^7$Be from $^{40}$Cl,$^{40}$Ar,$^{40}$Ca+$^{58}$Fe,$^{58}$Ni collisions at E/A=33 and 45 MeV showed a transition from a complete N/Z equilibration of the system (at E/A=33 MeV) to a non-equilibration (at E/A=45 MeV) where the reaction time is shorter and a strong memory of the N/Z in the initial interacting projectile and target is observed [22] in the decay channel. However this result is in contradiction to more recent observation that complete stopping is not achieved at E/A=30 MeV [23].

Stimulated by these results, we use isospin tracing observables to probe the isospin diffusion phenomenon occurring when two nuclei with different N/Z asymmetries stay in contact during the reaction. The isospin diffusion results from $,^{124}$Sn+$^{112,124}$Sn peripheral collisions at E/A=50 MeV and their rapidity dependence have provided one of the strongest probes of the density dependence of the symmetry energy in asymmetric nuclear matter [17]. (The rapidity $y = \frac{1}{2}\ln\left(\frac{E+p_{\parallel}c}{E-p_{\parallel}c}\right)$ reduces non-relativistically to $p_{\parallel}/(mc)$, where $E$, m and $p_{\parallel}$ are the total energy, mass and the momentum component parallel to the beam, respectively.) The short reaction time does not allow a complete N/Z equilibration of the system and a N/Z transparency is observed.

In this work we extend the previous isospin diffusion studies to lower beam energy of E/A=35 MeV. Regardless the longer interaction times between quasi-projectile and quasi-target, as compared to the same reaction system studied at E/A=50 MeV, the impact parameter and rapidity dependence of isospin tracing ratios show that the isospin degree of freedom remains non equilibrated even in the most dissipative central collisions. A comparison of the experimental results to quantum molecular dynamics calculations [24, 25] confirms the previously determined constraints on the density dependence of the symmetry energy [17].

The experiment was performed at the Laboratori Nazionali del Sud, Catania, Italy. $^{112}$Sn+$^{112}$Sn, $^{112}$Sn+$^{124}$Sn, $^{124}$Sn+$^{112}$Sn, $^{124}$Sn+$^{124}$Sn collisions were studied by bombarding $^{112}$Sn and $^{124}$Sn targets of 627μg/cm$^2$ and 689 μg/cm$^2$ areal density, respectively, with 35 MeV per nucleon $^{112}$Sn and $^{124}$Sn beams. Light charged particles (Z=1,2) and intermediate mass fragments (3≤Z≤10) were detected with the Chimera multidetector array, consisting of 1192 Silicon CsI(Tl) telescopes and subtending 94% of the total (4π) solid angle [26]. Impact parameter selection was provided by the charged particle multiplicity or the transverse energy of charged particles measured in the Chimera Array, both of which should decrease monotonically with impact parameter [27]. Under that assumption, one can define a reduced impact parameter based on the following relationship [27,28] (given here for a selection based on multiplicity)

$$\hat{b}(N_c) = \frac{b(N_c)}{b_{max}} = \left[\sum_{N_c}^{\infty} P(N_c)\right]^{1/2} / \left[\sum_{N_c(b_{max})}^{\infty} P(N_c)\right]^{1/2}, \quad (1)$$

where P($N_c$) is the probability distribution of the charge particle multiplicity $N_c$ corresponding to impact parameter, b. By setting selected experimental runs to $N_c(b_{max})$=1, we obtained values of $b_{max}$=8.8±1.5, 8.7±1.5, 8.4±1.5 and 8.6±1.5 fm, for reactions $^{112}$Sn+$^{112}$Sn, $^{112}$Sn+$^{124}$Sn, $^{124}$Sn+$^{112}$Sn, and $^{124}$Sn+$^{124}$Sn, respectively. The uncertainties here reflect uncertainties in the cross section for $N_c(b_{max})$≥1 stemming from uncertainties in the current integration. We note that analyses, using impact parameters derived from the transverse energy, $E_t$, or $N_c$, yield results that are indistinguishable.

The planar Si detectors ( ≈300 μm) of the Chimera telescopes used in these analyses were calibrated to about 2% using precision pulsers and the punch-through energies of alpha particles. The CsI(Tl) crystals of the telescope were calibrated for specific isotopes to an accuracy < 5% from the corresponding energy losses of these isotopes in the Si detectors. Planar silicon detectors and pulse shape discrimination in the CsI(Tl) crystals allowed isotopic resolution for $^7$Li, $^7$Be and other fragments. The isospin tracer technique of ref. [14, 29] probes N/Z stopping or transparency by rapidity dependence of an isospin transport ratio

$$R_i = \frac{2x_i - x_{A+A} - x_{B+B}}{x_{A+A} - x_{B+B}} \quad (2)$$

where $x$ is an isospin sensitive observable. Following ref. [19], we choose the observable, $x_7 = \ln(Y(^7Li)/Y(^7Be))$. Both experimental data and theoretical calculations predict $x_7$ to be a linear function of the asymmetry ($\delta = (\rho_n - \rho_p)/(\rho_n + \rho_p)$) of the emitting source. (Here, $\rho_n$ and $\rho_p$ denote the neutron and proton number densities respectively.) If $x_7$ depends linearly on $\delta$, then the isospin transport ratio calculated using $x_7$ equals the one calculated directly from the $\delta$ of the source [19], facilitating comparisons to transport model calculations.

In this work, we evaluate $R_i$ using values for $x$ obtained from symmetric systems A+A ≡ $^{124}Sn+^{124}Sn$ and B+B ≡ $^{112}Sn+^{112}Sn$, and asymmetric systems A+B (≡ $^{124}Sn+^{112}Sn$) and B+A (≡$^{112}Sn +^{124}Sn$). By construction, $R_i$ is automatically normalized to +1 and –1, for reactions i≡A+A and i≡B+B, respectively and in the limit of isospin equilibrium, $R_i=0$. In the limit of complete transparency, the observable $R_i$ should be +1 and -1 for the mixed systems.

The $^7Li$ and $^7Be$ yields were selected by rapidity and the associated charged particle multiplicity. The multiplicity was used to define a reduced impact parameter for each event following Eq. 1. Gaussian weighted sums of these yields over impact parameter,

$$\langle Y_i(y,\hat{b}) \rangle = \sum_{N_c} w(\hat{b}, N_C) Y_i(y, N_C) \quad (3)$$

using weights of the form

$$w(\hat{b}, N_c) = C \cdot \exp\left(-[\hat{b} - \hat{b}(N_c)]^2 / [2\sigma^2]\right) / y_{tot,Li+Be}(N_c), \quad (4)$$

with $\sigma = 0.4$ fm, were used to obtain average values of the $^7Li$ and $^7Be$ yields at selected reduced impact parameters of $\hat{b}$=0.12, 0.23, 0.34, 0.46, 0.58, 0.69, 0.81 and 0.93. (Eq. 1 contains an inverse weighting with respect to the total yields of $^7Li$ plus $^7Be$, which counters the strong dependence of the measurements upon $N_c$, and centers the contributions to these sums at the correct values for $\hat{b}$.)

Fig. 1 shows the center of mass energy spectra of the $^7$Li (open and closed circles) and $^7$Be (open and closed squares) for central ($\hat{b}$ <0.12; left panel) and mid-peripheral ($\hat{b}$ ~0.6; right panel) collisions. The solid symbols correspond to particles emitted in the neutron-rich $^{124}$Sn+$^{124}$Sn reactions while the open symbols correspond to particles emitted in the neutron-deficient $^{112}$Sn+$^{112}$Sn collisions. More neutron-rich $^7$Li particles are emitted during the $^{124}$Sn+$^{124}$Sn collisions and more neutron-deficient $^7$Be particles are emitted during the $^{112}$Sn+$^{112}$Sn collisions. Due to Coulomb effects, the shapes of the energy spectra for $^7$Li and $^7$Be are quite different. But the shapes of the energy spectra of the same isotope from both the $^{124}$Sn+$^{124}$Sn and $^{112}$Sn+$^{112}$Sn reactions are similar. To reduce statistical uncertainties and to explore the impact parameter dependence in greater precision, we compute the isospin diffusion ratios of Eq. 1 by combining the ratios obtained in rapidity regions placed symmetrically above and below the mid rapidity using the identities $x_{A+A}(y/y_{beam}) = x_{A+A}(0.5 - y/y_{beam})$ and $x_{B+B}(y/y_{beam}) = x_{B+B}(0.5 - y/y_{beam})$. Fig. 2 shows the resulting $R_7$, computed with $^7$Li and $^7$Be fragments emitted near projectile rapidity, $y \sim y_{beam}$, as a function of impact parameter. The displayed results show a trend of incomplete equilibration or incomplete stopping even at the smallest impact parameters. $R_7$ remains roughly independent of impact parameter at $\hat{b}$ < 0.6. At larger impact parameters $R_7$ increases with $\hat{b}$. Such observations indicate less mixing in peripheral collisions and more mixing in central collisions. However, even at the most central collisions, the magnitude of $R_7$ is far from zero, indicating that complete mixing and equilibration rarely occurs for the most central collision gates. The existence of very rare central events [30] where a complete N/Z mixing occurs cannot be excluded by our analysis. The resulting correlation between $R_7$ and impact parameter remains essentially the same if the transverse energy ($E_t$) is used instead of $N_C$ for impact parameter selection. Furthermore, our results support recent analyses of stopping in central Xe+Sn collisions over a large range of incident energies that suggest the non-attainment of equilibrium in central collisions and a minimum in the "stopping" at about 30-40 MeV per nucleon [23]. Since early works on the isospin degree of freedom in heavy-ion reactions, the neutron/proton N/Z-asymmetry was recognized as one of the most quickly evolving degrees of freedom [7-13]. Our findings about N/Z transparency in central collisions therefore suggest that even other degrees of freedom have not achieved their equilibrium values, with the consequence that the main assumptions taken for granted in statistical models of multifragmentation may not be satisfied. Microscopic transport models may therefore be more suited to study isospin diffusion and equilibration in heavy-ion collisions [12].

To compare our observations to transport theories, we calculated isospin transport ratios for this reaction with the Improved Quantum Molecular Dynamics (ImQMD) transport model of ref. [24]. Detailed description of the model and its application to the neutron/proton double ratio data and to isospin diffusion data can be found in refs. [17, 25]. Consistent with ref. [17, 25], we chose an Equation of State (EoS) with an isoscalar incompressibility of K=205 MeV and in-medium cross-sections that evolve to the free values at vanishing density [24], and explored the sensitivity of the isospin transport ratio to the density dependence of the symmetry energy, choosing the form

$$S(\rho) = \frac{C_{s,k}}{2}\left(\frac{\rho}{\rho_0}\right)^{2/3} + \frac{C_{s,p}}{2}\left(\frac{\rho}{\rho_0}\right)^{\gamma_i}. \qquad (5)$$

In our study, the kinetic and potential parameters are $C_{s,k}$=25 MeV, $C_{s,p}$=35.2 MeV and the symmetry energy at saturation density, $S_0 = S(\rho_0) = 30.1$ MeV.

Figure 3 compares the measured isospin transport ratios for $\tilde{b}$=0.1 (open circle), 0.23(open square), 0.34(open diamond), 0.46 (open triangle), 0.58 (open star) to predictions from the ImQMD models (lines). The left panel shows the comparison for a softer symmetry energy with $\gamma_i$=0.5 and the right panel shows the comparison for a stiffer symmetry energy with $\gamma_i$=2.0 assuming $b_{max}$ = 12 fm. Both calculations predict impact parameter independent values for the isospin transport ratios at b≤8 fm. Similar to the results of ref. [17], the isospin transport ratios are better reproduced by the softer symmetry energy with $\gamma_i$=0.5. We note that the value for $b_{max}$=12 fm used here reproduces the impact parameter dependence of $R_i$ rather well as shown in Fig. 2 but this value is larger than those estimated experimentally, albeit with large uncertainties. Changing the value of $b_{max}$ would not change the results at small impact parameters where both calculations and data are only weakly impact parameter dependent. Thus, the experiment confirms the transport theory predictions that stopping and chemical equilibrium are not the typical outcomes of a central heavy ion collision near the Fermi velocity.

In summary, we have studied the isospin equilibration in Sn+Sn collisions at incident energy of E/A=35 MeV using the isospin diffusion ratios constructed from the yields of $^7$Li and $^7$Be. The results, including the rapidity and impact parameter dependence of the isospin diffusion, agree with the transport model predictions based on the previously established symmetry energy constraints. Despite the longer reaction times, as compared to previous studies on the same system at E/A=50 MeV, we observe that complete equilibration is not achieved even for central collisions. Since the relaxation time of the isospin degree of freedom is expected to be very short, our results contradict the frequently used approximation that an equilibrated source is formed in the central collisions between heavy ions around Fermi energy.

**Figure Captions:**

**Figure 1:** (Color online) Left panel: Center of mass energy spectra of the $^7$Li (circles) and $^7$Be (squares) for central collisions of $^{124}$Sn+$^{124}$Sn reactions (closed symbols) and $^{112}$Sn+$^{112}$Sn reactions (open symbols). Right panel: Center of mass energy spectra of the $^7$Li and $^7$Be particles emitted in the peripheral Sn+Sn collisions.

**Figure 2:** (Color online) Isospin transport ratios constructed with the yield ratios of $^7$Li and $^7$Be particles emitted near projectile rapidity, y~$y_{beam}$, plotted as a function of the reduced impact parameters. The line is the result of ImQMD calculations using the softer symmetry potential with $\gamma_i$=0.5.

**Figure 3:** (Color online) Left panel: Isospin transport ratios plotted as a function of the normalized rapidity. The lines are ImQMD calculations using the softer symmetry potential with $\gamma_i$=0.5. Right panel: Same as left panel but with a stiffer symmetry potential of $\gamma_i$=2.0 is used in the calculation. The symbols in both panels are data with different reduced impact parameter cut. See text for details.

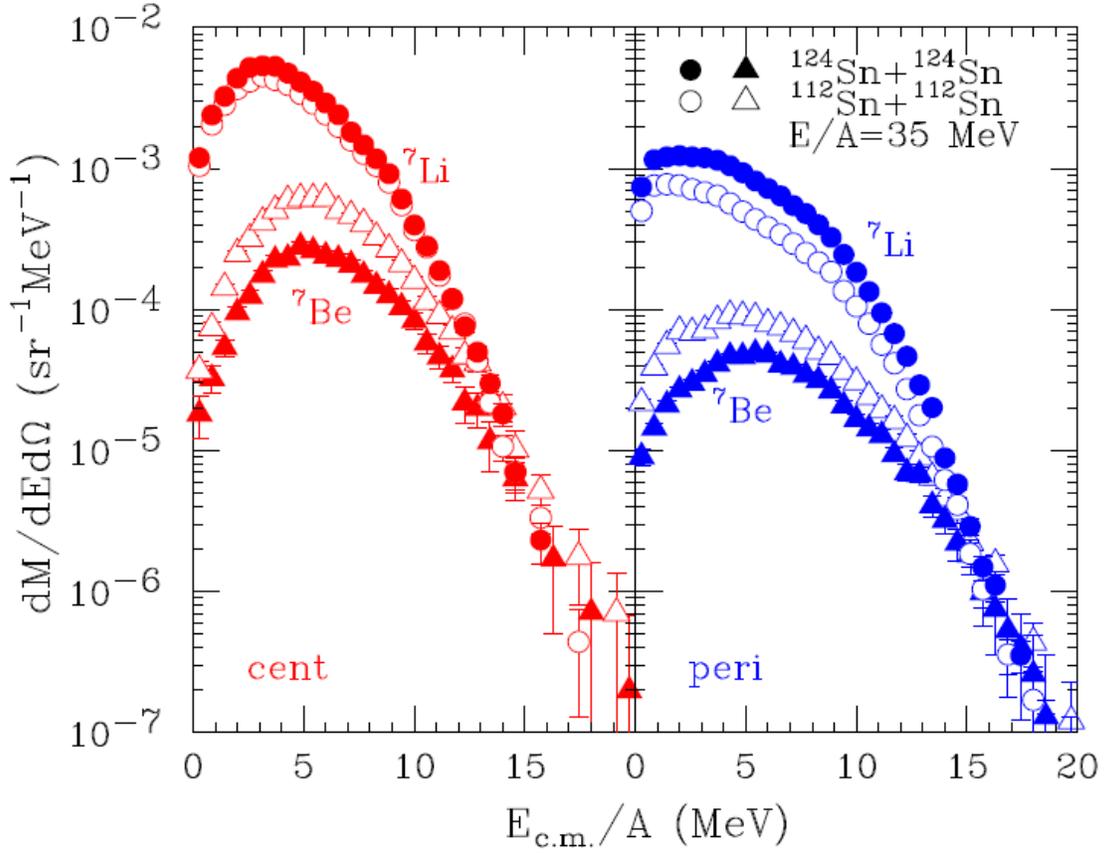

**Figure 1:** (Color online) Left panel: Center of mass energy spectra of the $^7$Li (circles) and $^7$Be (squares) for central collisions of $^{124}$Sn+$^{124}$Sn reactions (closed symbols) and $^{112}$Sn+$^{112}$Sn reactions (open symbols). Right panel: Center of mass energy spectra of the $^7$Li and $^7$Be particles emitted in the peripheral Sn+Sn collisions.

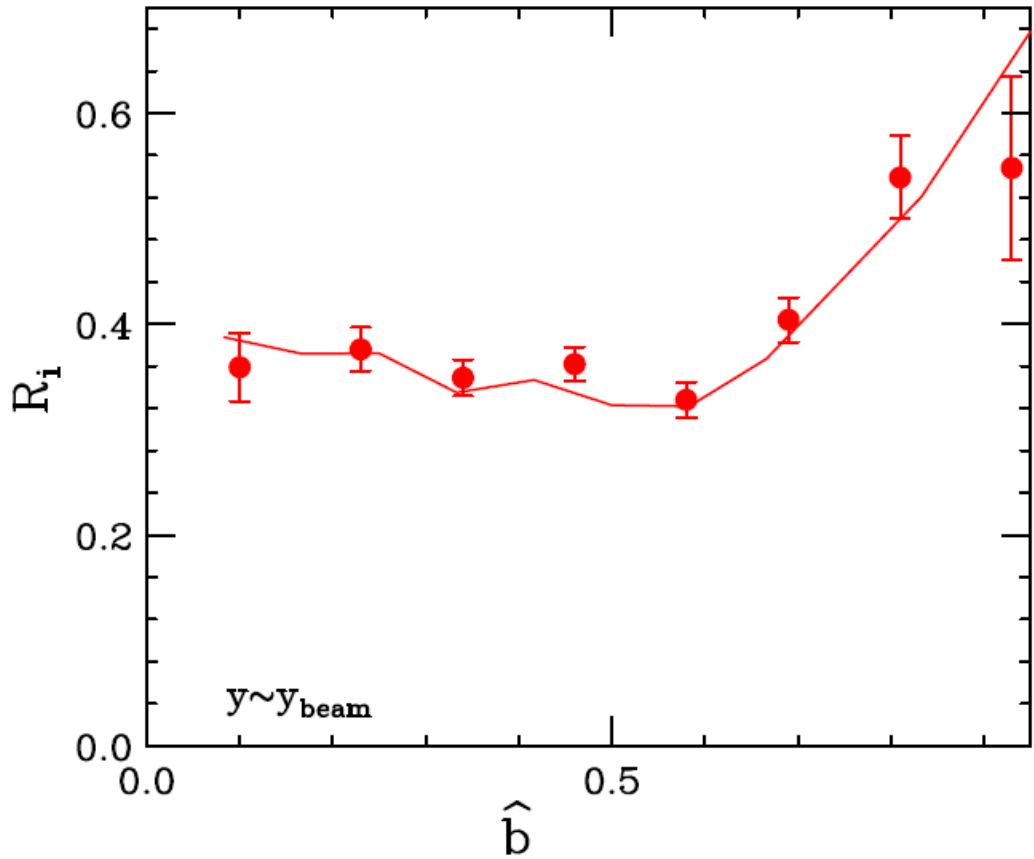

**Figure 2:** (Color online) Isospin transport ratios constructed with the yield ratios of $^7$Li and $^7$Be particles emitted near projectile rapidity, y~y$_{beam}$, plotted as a function of the reduced impact parameters. The line is the result of ImQMD calculations using the softer symmetry potential with $\gamma_i$=0.5.

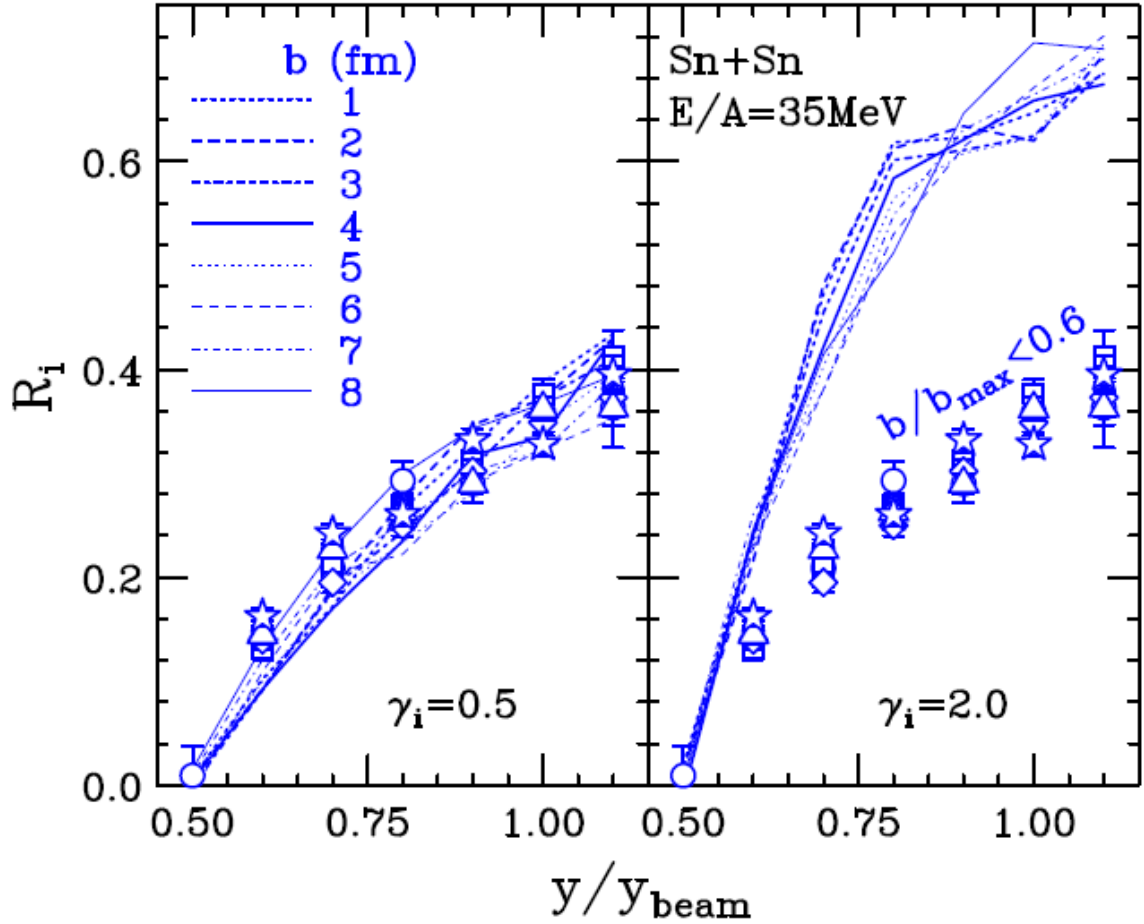

**Figure 3:** (Color online) Left panel: Isospin transport ratios plotted as a function of the normalized rapidity. The lines are ImQMD calculations using the softer symmetry potential with $\gamma_i$=0.5. Right panel: Same as left panel but with a stiffer symmetry potential of $\gamma_i$=2.0 is used in the calculation. The symbols in both panels are data with different reduced impact parameter cut. See text for details.